\documentclass[11pt,a4paper]{article}
\pdfoutput=1
\usepackage{jheppub}
\usepackage{etoolbox}

\def\one{{\,\hbox{1\kern-.8mm l}}}
\newcommand{\Dslash}{\not{\hbox{\kern-4pt $D$}}}
\newcommand{\pdslash}{\not{\hbox{\kern-2pt $\partial$}}}
\newcommand{\cL}{\mathcal{L}} 
\newcommand{\bcL}{\mathcal{\bar L}}

\newcommand{\Comment}[1]{{}}

\allowdisplaybreaks

\def\IZ{{\mathbb Z}}

\def\IR{{\mathbb R}}

\newcommand{\bc}{\begin{center}}
\newcommand{\ec}{\end{center}}
\newcommand{\ba}{\begin{array}}
\newcommand{\ea}{\end{array}}
\newcommand{\beq}{\begin{equation}}
\newcommand{\eeq}{\end{equation}}
\newcommand{\bea}{\begin{eqnarray}}
\newcommand{\eea}{\end{eqnarray}}
\newcommand{\bmx}{\begin{pmatrix}}
\newcommand{\emx}{\end{pmatrix}}
\newcommand{\nn}{\nonumber}

\newcommand{\be}{\begin{equation}}
\newcommand{\ee}{\end{equation}}

\newcommand{\del}{\partial}
\newcommand{\half}{{\frac{1}{2}\,}}

\newcommand{\cbar}{{\bar c}}

\newcommand{\tT}{{\tilde T}}

\def\IB{\relax{\rm I\kern-.18em B}}
\def\IC{{\relax\hbox{\kern.3em{\cmss I}$\kern-.4em{\rm C}$}}}
\def\ID{\relax{\rm I\kern-.18em D}}
\def\IE{\relax{\rm I\kern-.18em E}}
\def\IF{\relax{\rm I\kern-.18em F}}
\def\II{\relax{\rm I\kern-.18em I}}
\def\IZ{\relax{\sf Z\kern-.35em Z}}
\def\Id{\relax{1\kern-.32em 1}}
\def\IG{\relax\hbox{$\inbar\kern-.3em{\rm G}$}}
\def\IR{\relax{\rm I\kern-.18em R}}
\newcommand\sfrac[2]{{\textstyle\frac{#1}{#2}}}

\newcommand\shalf{{\textstyle\frac12}}

\title{Free-field realisations of the BMS$_3$ algebra and its extensions}

\author{Nabamita Banerjee$^*$, Dileep P. Jatkar$^\dagger$, Sunil
  Mukhi$^*$ and Turmoli Neogi$^*$} 
\author{}

\affiliation{$^*$Indian Institute of Science Education and Research,\\
Homi Bhabha Rd, Pashan, Pune 411 008, India}
\affiliation{$^\dagger$Harish-Chandra Research Institute,\\  Chhatnag
  Road, Jhunsi, Allahabad 211019, India}

\emailAdd{nabamita@iiserpune.ac.in, dileep@hri.res.in,
  sunil.mukhi@iiserpune.ac.in, turmoli.neogi@students.iiserpune.ac.in}

\abstract{We construct an explicit realisation of the BMS$_3$ algebra
  with nonzero central charges using holomorphic free fields. This can
  be extended by the addition of chiral matter to a realisation having
  arbitrary values for the two independent central charges. Via the introduction of additional free fields, we extend our construction to  the minimally supersymmetric BMS$_3$ algebra and to
the nonlinear higher-spin BMS$_3$-W$_3$ algebra. We also describe an extended system that realises both the SU(2) current algebra as well as BMS$_3$ via the Wakimoto representation, though in this case introducing a central extension also brings in new non-central operators.}

\preprint{}

\keywords{BMS symmetry, Flat holography, Conformal field theory}

\makeatletter
    \patchcmd{\maketitle}{\@fpheader}{\includegraphics[height=15mm]{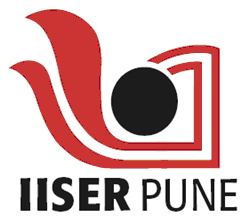}}{}{}
    \makeatother
    
\begin{document}

\maketitle

\section{Introduction}

The first concrete realisation of gauge-gravity holography was
obtained using D-brane configurations in string theory and is embodied
in the famous AdS/CFT correspondence\cite{Maldacena:1997re}.  While
AdS/CFT provides very explicit examples, the concept of holography
originated in the more general context of gravitational physics and
black holes\cite{Hooft:1993gx,Susskind:1994vu} without any restriction
to AdS or asymptotically AdS space-time.  Thus it remains an important
goal to understand holography in theories with flat or deSitter
asymptotics.

The AdS/CFT correspondence has taught us how to build a dictionary
which relates a theory of gravity with its dual non-gravitational
theory.  One of the insights obtained from this
duality\cite{Brown:1986nw} is that the asymptotic symmetry algebra in
the gravity theory on a manifold with boundary gives the symmetry of
the dual non-gravitational theory on that boundary. Thus one expects
the asymptotic symmetry of the bulk theory with flat or deSitter
asymptotics to hold the key to the extension of holography beyond the
AdS context. In this work we focus on the flat space case.

It has been known for quite some time that the asymptotic symmetry
algebra on $\mathcal{I}^+$ for gravity theories with flat-space
asymptotics is given by the Bondi-van der Burg-Metzner-Sachs (BMS)
algebra\cite{Bondi:1962px,Sachs:1962zza}.  It is natural to ask if
this algebra encodes information about bulk gravity with flat space
asymptotics. Several interesting works have shed some light on this
question\cite{Barnich:2010eb,Bagchi:2010eg,Bagchi:2010zz,Bagchi:2013qva,Schulgin:2013xya,Barnich:2014kra,Barnich:2015uva}. One
outcome of these investigations is a detailed formulation of the
infinite-dimensional BMS algebras in both 3+1 and 2+1 dimensions. In
the latter case, the relevant algebra is known as BMS$_3$, and is an
extension of a single Virasoro algebra by an additional chiral spin-2
generator along with two independent central charges.

It is interesting to ask if there are concrete realisations of the BMS
algebra in terms of quantum fields. Besides the motivation of flat-holography, this could be interesting just from the point of view of conformal field theory. In principle realisations can be studied in any dimension but since the BMS$_3$ case has a Virasoro algebra as a sub-algebra, one may expect that the language of conformal field
theory will be helpful in finding classes of realisations. Moreover,
three dimensional gravity is relatively simple in the bulk and
two-dimensional field theories (conformal and otherwise) have been
studied for many decades. Therefore, understanding BMS$_3$/FT$_2$
seems like an ideal place to start. 

There already exist several works attempting to describe the boundary
theory for gravity/supergravity in an asymptotically flat bulk
\cite{Barnich:2013yka,Schulgin:2013xya,Barnich:2015sca}. These duals
are rather complicated and in some cases involve non-polynomial
constructions in terms of fields. In this note we will approach the
problem from the opposite direction: we construct a very simple and
explicit free-field realisation of BMS$_3$. Remarkably it turns out
that this realisation provides non-vanishing central charges $c_1,c_2$
for BMS$_3$, of which $c_1$ is fixed. To realise an arbitrary  $c_1$,
it is sufficient to adjoin arbitrary chiral matter to this system. This freedom can be used in particular to set $c_1=0$, which is the value for pure Einstein gravity. However our construction can also address more general asymptotically flat bulk theories with nonzero $c_1$. The important thing is that we achieve $c_2\ne 0$, which is the case in typical bulk  realisations of BMS$_3$ including Einstein gravity. The value of $c_2$ per se is not important as it can be changed by simply rescaling the $M_n$ generators of BMS$_3$. However whether it is zero or non-zero is significant and our construction successfully permits the latter choice. It must be pointed out, though, that there are some obstacles in going from our present construction to an asymptotically flat gravitational dual for a reasons spelt out below, and in the present work we will not attempt this.

Before describing our construction, we would like to motivate the relevance of free-field representations of two-dimensional conformal algebras, known to mathematicians as infinite-dimensional Fock space representations, with some historical background. Free fermion and boson representations of affine Lie algebras originated in work of Bardakci and Halpern\cite{Bardakci:1970nb,Halpern:1975nm}. These ideas and subsequent developments were then rigorously systematised by mathematicians\cite{Lepowsky:1978jk,Frenkel:1980rn,Segal:1981ap} and the results had a remarkable impact on the role of affine Lie algebras in both mathematics and physics. Among other things they led to the vertex-operator construction of the Monster algebra\cite{Frenkel:book}, as well as the understanding of enhanced gauge symmetry in toroidal compactifications of string theory\cite{Narain:1985jj}. Some years later Wakimoto made the important discovery of free-field realisations of SU(2) at arbitrary level\cite{Wakimoto:1986gf} and this in turn was generalised to SU(n) after considerable effort by different groups (for a nice review, see Ref.\cite{Bouwknegt:1990wa}). Free-field realisations of coset and W-algebras were discovered during the 1990's\cite{Bouwknegt:1992wg, Bouwknegt:1995ag}. This background should make it clear that there is no straightforward recipe to find free-field realisations of an arbitrary conformal algebra, and also that when such a realisation exists it is usually of significance.

Returning now to a summary of our construction, it starts with a chiral
$\beta$-$\gamma$ bosonic ghost system with spins $(2,-1)$. A twist of
the energy-momentum tensor\footnote{To avoid potential confusion, we
  should emphasise that we are using language associated to chiral
  conformal theories, but this should not be taken too literally. For
  example what we call the ``energy-momentum tensor'' does not
  literally have to be the energy-momentum tensor of a boundary theory
  on ${\cal I}^+$.} by adding a total derivative term leads to the
BMS$_3$ algebra complete with nonzero central extensions $c_1,c_2$. Notably, our twist is {\em not} by the derivative of a primary current, as has usually been the case in 2d topological field theory and related contexts\cite{Knizhnik:1988ak,Eguchi:1990vz,Mukhi:1993zb}. Despite this, the twist preserves the Virasoro algebra and we find that the Virasoro central charge (denoted below by $c_1$) is equal to 26. To achieve a general value of $c_1$,
we add chiral matter with central charge $c_0$ and verify that the
algebra is completely unchanged by this addition, except that
$c_1$ is modified to $c_0+26$. A suitable choice of $c_0$
can thus realise any positive or negative value of $c_1$. Meanwhile
the other central charge (denoted $c_2$ below) is proportional to the
twist, so in the end we have arbitrary central charges $c_1,c_2$ for
the BMS$_3$ algebra.
 
In our second example we start with the Wakimoto free field
representation\cite{Wakimoto:1986gf} of a chiral affine SU(2) algebra.
The energy-momentum tensor obtained from the Sugawara construction is
twisted by the divergence of the third component of the SU(2) current,
and then deformed further by adding a total derivative term, to obtain
a realisation of a coupled SU(2)-BMS$_3$ algebra.  Our third example is a
supersymmetric version of BMS$_3$ algebra. For this we introduce another pair of
ghosts $b,c$ that are fermionic and have spins
$(\frac32,-\frac12)$. Choosing the super-charge $Q$ to be an appropriate
function of $b,\beta$ and $c$, we realise super-BMS$_3$ with again two
independent central charges. Finally, we extend this construction to 
the $W_3$ case of higher-spin-BMS$_3$. 

At each step we need to introduce some new fields as well as new ways
of using them. It is remarkable that, despite the system being highly
overconstrained, we are able to precisely realise the desired algebra
every time. It is worth mentioning that in the case of BMS-W$_3$ the
normal-ordering prescription for the nonlinear terms is extremely
subtle, and great care is needed in verifying that the algebra is
correctly realised.

It is important to mention that for all of our constructions, we have chosen a vacuum defined by the highest weight condition:
\begin{equation}
L_n |0\rangle = M_n |0\rangle =0 , \qquad n\geq -1
\end{equation}
where $L_n$ and $M_n$ are modes of the BMS$_3$ algebra in Eq.(\ref{eq:11}), as well as analogous conditions for the additional modes appearing in the supersymmetric and $W_3$ cases. This is the choice made in Ref.\cite{Grumiller:2014lna}. As pointed out there, with this choice one does not get unitary representations of BMS$_3$ (when $c_2>0$). Moreover these conditions violate parity. These are potentially serious issue for applying the construction to flat-space holography. Alternative norms have been discussed in the literature \cite{Campoleoni:2016vsh}. It will be important to examine these and more general possibilities in the context of free-field realisations before these can be applied in the  holographic context. As mentioned above, this construction may be of field-theoretic interest even outside the context of holography.

\section{Free fields and BMS$_3$}
\label{sec:free-fields-bms_3}

The BMS$_3$ algebra can be derived in various ways\cite{Sachs:1962zza,Bondi:1962px,Barnich:2006av,Barnich:2010eb}. One relatively simple way is to start with the asymptotic symmetry algebra corresponding to asymptotically AdS$_3$ spacetime. It is a celebrated result\cite{Brown:1986nw} that this is a pair of 
independent Virasoro algebras:
\begin{equation}
  \label{eq:10} 
\begin{split}
[\cL_n,\cL_m]&=(n-m)\cL_{n+m} +
\frac{c}{12}\,n(n^2-1)\,\delta_{n+m,0}\ ,\\
[\bcL_n,\bcL_m]&=(n-m)\bcL_{n+m} +
\frac{\cbar}{12}\,n(n^2-1)\,\delta_{n+m,0}\ ,
\end{split}
\end{equation}
with, in principle, independent central charges $c,\cbar$. If one specialises to ordinary gravity at two-derivative order one finds $c={\bar c}=\frac{3\ell}{2G}$. The BMS$_3$ algebra arises in the limit where one takes the AdS$_3$ radius $\ell$ to infinity. This has to be done by scaling the generators of the two Virasoro algebras carefully while taking the limit. For this, one takes the linear combinations:
\be
L_n={\cL_n-\bcL_{-n}},\quad M_n=\epsilon(\cL_n+\bcL_{-n})
\ee
in the limit $\epsilon\to 0$. One finds that $c_1=c-\cbar$ and
$c_2=\epsilon(c+\cbar)$, hence to get a finite $c_1,c_2$ from this
starting point we take $c,\cbar\to\infty$ keeping $c-\cbar$
fixed. The resulting algebra in this limit, known as BMS$_3$, is:
\begin{equation}
  \label{eq:11}
  \begin{split}
[L_n,L_m]&=(n-m)L_{n+m}+\frac{c_1}{12}\,n(n^2-1)\,\delta_{n+m,0}\\
[L_n,M_m]&=(n-m)M_{n+m}+\frac{c_2}{12}\,n(n^2-1)\,\delta_{n+m,0}\\
[M_n,M_m] &=0\ .
\end{split}
\end{equation}

As explained in the introduction, we are interested in finding an infinite-dimensional Fock-space representation of the above algebra. For convenience we introduce a holomorphic coordinate $z$ (as in the
previous footnote, we emphasise that this should not literally be taken as the
coordinate of an underlying space or space-time) and construct
canonical fields:
\be
\begin{split}
T(z) &= \sum_{n\in \IZ} L_n z^{-n-2}\\
M(z) & = \sum_{n\in \IZ} M_n z^{-n-2}
\end{split}
\ee
We are going to construct the two fields $T(z),M(z)$ in terms of holomorphic free fields. It is important to keep in mind that the above is only a technical device. In principle one could keep working with the modes $L_n,M_n$ and construct them in terms of infinitely many pairs of modes satisfying canonical commutation relations, but the holomorphic approach has been recognised since the seminal work of Belavin, Polyakov and Zamolodchikov\cite{Belavin:1984vu} to be more practical. 

Using the relation between algebras and operator product expansions we
can rewrite the BMS$_3$ algebra in terms of the operator product
expansion of the fields: 
\begin{equation}
  \label{eq:12}
  \begin{split}
T(z)T(w)&\sim \half\frac{c_1}{(z-w)^4}+ \frac{2T(w)}{(z-w)^2}+
\frac{\del T(w)}{z-w}\\ 
T(z)M(w)&\sim \half\frac{c_2}{(z-w)^4}+ \frac{2M(w)}{(z-w)^2}+
\frac{\del M(w)}{z-w}\\
M(z) M(w)&\sim 0\ ,
\end{split}
\end{equation}
where, as usual, we write only the singular terms on the RHS.  The
field $T$ generates a Virasoro sub-algebra. The operator product
expansion between $T$ and $M$ tells us that $M$ is a dimension 2
field under the Virasoro algebra, but due to the presence of a central term it fails to be a primary field.

To obtain a free field representation of the BMS$_3$ algebra above, we start with a 
bosonic $\beta$-$\gamma$ satisfying the operator product expansion:
\begin{equation}
  \label{eq:14}
  \gamma(z)\beta(w)\sim \frac{1}{z-w} 
\end{equation}
Such a system has played an important role in many areas of conformal field theory and string theory, and a review of some aspects can be found in Ref.\cite{Nekrasov:2005wg}. One can take the conformal dimensions of $(\beta,\gamma)$ to be $(p,1-p)$ for any integer $p$. This is achieved by starting with the ``basic'' pair with dimensions $(1,0)$ and twisting the energy-momentum tensor suitably by a derivative of the ghost-number current $:\!\beta\gamma\!:$. 
For our purposes, we would like $(\beta,\gamma)$ to have dimensions $(2,-1)$ and will work with an energy-momentum tensor that has already been twisted to achieve this, which turns out to be: 
\be
T_{\beta,\gamma}=- 2\!:\!\beta\del\gamma\!:-:\!\gamma\del\beta\!: \ .
\ee
It immediately follows that
\begin{equation}
  \label{eq:15}
  \begin{split}
T_{\beta,\gamma}(z) \beta(w)&\sim \frac{2\beta(w)}{(z-w)^2}+\frac{\del
    \beta(w)}{z-w} \\
T_{\beta,\gamma}(z) \gamma(w)&\sim \frac{-\gamma(w)}{(z-w)^2}+\frac{\del
    \gamma(w)}{z-w}\ .
\end{split} 
\end{equation}
As is familiar, the central charge of $T_{\beta,\gamma}$ is $26$.  

Let us note here that the pair of spin-2 fields $(T_{\beta,\gamma}(z),\beta(z))$ generate something close to the BMS$_3$ algebra if we identify them with the BMS$_3$ generators $(T(z),M(z))$. Indeed the OPE of $T_{\beta,\gamma}$ with itself, together with the second equation above, form a BMS$_3$ algebra with $c_1=26,c_2=0$. As we will see, the latter is the bigger problem -- a general BMS$_3$ requires nonvanishing $c_2$. This does not arise in the present system because $\beta$ is primary.

To remedy this problem and introduce $c_2\ne 0$, we twist the energy-momentum tensor:
\begin{equation}
  \label{eq:16}
  T(z) = T_{\beta,\gamma} - a\,\del^3\gamma 
\end{equation}
where $a$ is an arbitrary constant. As mentioned in the introduction, this twist is not of the form $T(z)\to T(z)+\half\del J(z)$ for a primary current $J(z)$. Indeed in the present case, $J(z)$ would be proportional to $\del^2\gamma$ which is definitely not primary, being the descendant of the primary $\gamma$ under the action of ${\cal L}_{-1}^2$. As a result it is not at all clear that the above twist preserves the Virasoro algebra. What is evident, however, is that it will induce a fourth-order pole, or in other words a central term, in the $T(z)\beta(w)$ OPE. 

The calculation of the OPE of $T(z)$ above with itself involves cross terms between $T_{\beta,\gamma}$ and $\del^3\gamma$. At intermediate stages these give rise to poles of up to fifth-order. For $T(z)$ to continue to satisfy the Virasoro algebra, the fifth- and third-order poles must cancel completely while the second and first-order poles should depend only on $T(z)$ and not separately on $\del^3\gamma$. None of these facts can be influenced by a choice of the coefficient $a$ (since all the cross terms are proportional to $a$) and therefore the system is quite overdetermined. However a moderately tedious calculation shows that all the unwanted terms indeed cancel out and one gets:
\begin{equation}
  \label{eq:17}
  T(z)T(w)\sim
\half\frac{26}{(z-w)^4}+\frac{2\,T(w)}{(z-w)^2}+\frac{\del
  T(w)}{z-w}.
\end{equation}
Surprisingly, the new term $\del^3\gamma$ in $T$ has not
even changed the central charge, nor has it introduced any extra poles in
the $T$-$T$ OPE.  Instead, it just modifies the right-hand side of
(\ref{eq:17}) such that the final expression is again expressed in
terms of $T$.

Now choosing $M(z)=\beta(z)$, we see that the operator product expansion of
$\beta$ with $T$ is modified due to the twisting term, with the result:
\begin{equation}
  \label{eq:18}
  T(z)M(w) \sim  \half \frac{12a}{(z-w)^4} +
  \frac{2M(w)}{(z-w)^2}+\frac{\del M(w)}{z-w}
\end{equation}
Finally, because of the first-order nature of the ghost system we have:
\be
\label{eq:18a}
M(z)M(w)\sim 0
\ee

Combining the results in eq.(\ref{eq:17}) and eq.(\ref{eq:18}) we see
that together $T(z)$ and $M(z)$ define a BMS$_3$ algebra with
$c_1=26$ and $c_2=12a$. 
The central charge $c_2$ can be set to any nonzero value by tuning $a$. 
This freedom corresponds to the
fact that within the BMS$_3$ algebra one can always change the value
of $c_2$ by scaling $M$.

This construction lands us in a fixed value of the Virasoro central charge, namely,
$c_1=26$.  However, we can always couple any chiral conformal field
theory, whose energy momentum tensor $T_{\rm matter}$ has
central charge $c_0$, to this system.  The total energy-momentum tensor
will then be
\begin{equation}
  \label{eq:7}
  T(z) = T_{\rm matter}+T_{\beta,\gamma} - a\,\del^3\gamma\ ,
\end{equation}
resulting in a total central charge $c_1= c_0+26$ and no change in
any other operator product expansions. To summarise, we see that a
$(\beta,\gamma)$ system of spin $(2,-1)$, suitably combined with any
standard set of fields realising one copy of the Virasoro algebra,
gives an explicit realisation of the BMS$_3$ algebra with completely
arbitrary central charges.

\section{Free field representation of super-BMS$_3$}
\label{sec:free-field-repr}

In this section we will consider the minimal supersymmetric generalisation of
the BMS$_3$ algebra.  This is obtained by adjoining a single set of spin-$\frac32$ generators $Q_r$ to the BMS$_3$ algebra, with commutation relations\cite{Barnich:2014cwa, Barnich:2015sca}:
\begin{equation}
  \begin{split}
[L_n,L_m]&=(n-m)L_{n+m}+\frac{c_1}{12}\,n(n^2-1)\,\delta_{n+m,0}\\
[L_n,M_m]&=(n-m)M_{n+m}+\frac{c_2}{12}\,n(n^2-1)\,\delta_{n+m,0}\\
[M_n,M_m] &=0  \\
[L_n,Q_r] & =\left(\frac{n}{2}-m\right)Q_{n+r}  \\
[M_n,Q_r] &= 0\\
\{Q_s,Q_r\} &= M_{r+s}+\frac{c_2}{6}\,\left(s^2-\frac{1}{4}\right)\delta_{r+s,0}\ .
\end{split}
\end{equation}
Here $(r,s)$ are both integer or both half-integer.  For definiteness we will take it to be half-integer. This algebra can
be written in terms of the operator product as
\begin{equation}
  \label{eq:9}
  \begin{split}
        T(z)T(w) & \sim\ \half\frac{c_1}{(z-w)^4} + \frac{2T(w)}{(z-w)^2} +
    \frac{\partial T(w)}{z-w}, \\
    T(z)M(w) & \sim\  \half\frac{c_2}{(z-w)^4} + \frac{2M(w)}{(z-w)^2} +
    \frac{\partial M(w)}{z-w}, \\ 
    T(z)Q(w) & \sim\ \frac{\frac32 Q(w)}{(z-w)^2} + \frac{\partial
      Q(w)}{z-w}, \\
    Q(z)Q(w) & \sim\ \frac{1}{3}\frac{c_2}{(z-w)^3} + \frac{M(w)}{z-w}.
  \end{split}
\end{equation}
with the remaining OPE's being non-singular. We have supplemented $T(z),M(z)$ with the chiral field
$Q(z)=\sum_{r}Q_r z^{-r-\frac32}$. 

We would now like to obtain a free-field realisation of this algebra. For this, we supplement the
$\beta$-$\gamma$ system of section \ref{sec:free-fields-bms_3} by a
Grassmann-odd $b$-$c$ ghost system of spins
$\left(\frac32,-\frac12\right)$.  The OPEs of the $b$-$c$ fields are
given by
\begin{equation}
  \label{eq:22}
  b(z)c(w)\ \sim\ \frac{1}{z-w},\quad b(z)b(w)\ \sim \ 0,\quad
  c(z)c(w)\ \sim \ 0. 
\end{equation}
This system is well-known to have central charge $-15$. Now we proceed by choosing $T(z)$ to be the standard energy-momentum tensor for the fields $(\beta,\gamma)$ and $(b,c)$ of dimensions $(2,-1)$ and $(\sfrac32,-\shalf)$ respectively. Next we twist it by $-a\,\del^3\gamma$ as in the previous section. Letting $M(z)=\beta(z)$ as before, we obtain the bosonic part of the super-BMS$_3$ algebra with central charges $c_1=26-15=11$ and $c_2=12a$.

The next step is to represent the supersymmetry generator $Q(z)$. Since this has dimension $\frac32$, it would be most natural to start by realising it as $b(z)$. However, this would give rise to the OPE $Q(z)Q(w)\sim 0$ which is not what we want. So we try to supplement $b(z)$ by terms of the same dimension, namely $\frac32$, such that both $M(z)$ and a central term appear on the RHS of $Q(z)Q(w)$. Since $M(z)=\beta(z)$, the former requirement is achieved by adding a term of the form $\beta c$ in $Q$. For the central term, we must add a term proportional to $\del^2 c$. In this way we ensure that the desired terms arise, but in principle we might get several additional terms from the square of individual terms in $Q$ as well as from an additional cross term between $\beta c$ and $\del^2 c$. Fortunately every one of these terms vanishes. The squares of individual terms do not contribute to the singular OPE because each term contains only one of a pair of canonically conjugate variables. Similarly the additional cross term vanishes. Now the coefficients of the terms in $Q$ are easily adjusted to give the correct $Q(z)Q(w)$ OPE, and we find:
\be
Q(z)=b(z) + \half\!\!:\!\beta c\!:\!(z)+a\,\del^2 c(z) 
\ee 

At this stage, we have completely specified all the generators for the super-BMS$_3$ algebra:
\begin{equation}
  \label{eq:21}
  \begin{split}
    T(z) &= - \frac32 :\!b\partial c\!:\!(z) + \frac12 :\! c\partial b\!:\!(z) -
    2:\!\beta\del\gamma\!:\!(z)-:\!\gamma\del\beta\!:\!(z) - 
    a\, \partial^3\gamma(z), \\
    M(z) &= \beta(z), \\
    Q(z) &= b(z) + \frac12\! :\!\beta c\!:\!(z) + a\, \partial^2 c(z),
  \end{split}
\end{equation}
where $T(z)$ and $M(z)$ generate the three dimensional BMS algebra and
$Q(z)$ is the supersymmetry current. However we now have another potential problem. The OPE's $T(z)Q(w)$ and $M(z)Q(w)$ have not yet been verified, and there is no more freedom to adjust any of the generators. 

The first of these OPE's simply says that $Q(z)$ is a primary under $T(z)$. However, this seems impossible to ensure since, out of the two terms added to $b(z)$ to form $Q$, the term $\beta c$ appears to be primary because it is the product of two commuting free fields, while the term $\del^2 c$ is certainly not primary, being the second derivative of a primary. However one notices that the twist in $T$ by $\del^3\gamma$ renders the $\beta c$ term in $Q$ non-primary as well. Moreover, for the pre-determined values of the
coefficients of these terms, these two non-primary contributions neatly cancel! As a result we
find the desired OPE's:
\begin{equation}
  \label{eq:23}
  \begin{split}
    T(z)T(w) & \sim\ \frac{\frac{15}{2}}{(z-w)^4} + \frac{2T(w)}{(z-w)^2} +
    \frac{\partial T(w)}{z-w}, \\
    T(z)M(w) & \sim\  \frac{6a}{(z-w)^4} + \frac{2M(w)}{(z-w)^2} +
    \frac{\partial M(w)}{z-w}, \\ 
    T(z)Q(w) & \sim\ \frac{\frac32 Q(w)}{(z-w)^2} + \frac{\partial
      Q(w)}{z-w}, \\
    Q(z)Q(w) & \sim\ \frac{4a}{(z-w)^3} + \frac{M(w)}{z-w}.
  \end{split}
\end{equation}
Finally the OPE $M(z)Q(w)$ follows from the fact that $Q$ is independent of $\gamma(z)$. 
We have thus obtained a free-field representation of the super-BMS$_3$ algebra with
central charges $c_1=15$ and $c_2=12a$.

We have seen that the OPE of the supersymmetry current with itself in the minimal super-BMS$_3$ algebra considered in this section does not give $T(z)$, rather it produces $M(z)$. In this fact it differs significantly from the OPE of a supersymmetry charge in a standard (chiral) superconformal algebra, where $Q(z)Q(w)$ gives rise to $T(z)$. 
Hence if we want to augment our construction with other degrees of freedom to make the central charge $c_1$ arbitrary, we cannot achieve this by coupling the above system to a
superconformal field theory. Rather, we need to couple it to a bosonic conformal
field theory of chiral matter with central charge $c_0$. In this way only $T(z)$ changes while $M$ and $Q$ remain as before. Then we find an arbitrary value  $c_1=c_0+15$ for the first central charge of super-BMS$_3$. As before, the second central charge $c_2$ is proportional to a free parameter $a$ and was therefore arbitrary to start with.

\section{Spin 3 BMS$_3$ algebra}

We now attempt to find a free field representation of the $W_3$ BMS$_3$
algebra. In terms of modes, we have $L_n$ and $M_n$ as in the ordinary BMS$_3$ algebra, supplemented by $W_n$ and $V_n$. The algebra of these modes is given by the commutators: 
\cite{Bouwknegt:1992wg,Afshar:2013vka}\footnote{Our
  normalization of this equation is in agreement with Ref.\cite{Bouwknegt:1992wg} and differs from Ref.\cite{Afshar:2013vka} by a factor of 30 in the last two commutators.}:
\begin{align}
      [L_m,L_n] &= (m-n) L_{m+n} + \frac{c_1}{12}(m^3-m)\delta_{m,-n} \nn\\
    [L_m,M_n] &= (m-n) M_{m+n} + \frac{c_2}{12}(m^3-m)\delta_{m,-n} \nn\\
    [L_m,W_n] &= (2m-n) W_{m+n},\quad
    [L_m,V_n] = (2m-n) V_{m+n}\nn\\
    [M_m,W_n] &= (2m-n) V_{m+n}\nn\\
    [W_m,W_n] &=\frac{1}{30}\bigg[ (m-n)(2m^2 + 2n^2-mn-8)L_{m+n} \nn\\
    &\quad+ \frac{192}{c_2}(m-n)\Lambda_{m+n} -
    \frac{96(c_1+\frac{44}{5})}{c_2^2}(m-n)\Theta_{m+n} 
   \nn \\
    &\quad+
    \frac{c_1}{12}m(m^2-1) (m^2-4)\delta_{m,-n}\bigg]\nn\\
    [W_m,V_n] &= \frac{1}{30}\bigg[ (m-n) (2m^2 + 2n^2-mn-8)M_{m+n} \nn\\
    &\quad +\frac{96}{c_2}(m-n)\Theta_{m+n} 
    +\frac{c_2}{12}m(m^2-1) (m^2-4)\delta_{m,-n}\bigg],\label{eq:24}
  \end{align}
where $\Theta_m \equiv \sum_nM_nM_{m-n}$ and $\Lambda_m \equiv
\sum_n:\!L_nM_{m-n} \!:-\frac{3}{10}(m+2)(m+3)M_m$. 
Observe that $\Lambda$ contains a bilinear of two non-commuting modes $L$ and $M$ and therefore normal-ordering is necessary. It is also noteworthy that $\Lambda$ contains a term linear in $M$. 

To find a free field representation, note that upon conversion to fields, the generators of the above algebra are the pair of spin-2 fields $T(z)$ and $M(z)$ augmented by the pair of spin-3 fields $W(z)$ and $V(z)$. There are similarities between $M(z)$ and $V(z)$ in that both are self-commuting (viewed as modes) or equivalently have non-singular OPE's with themselves (as fields). 

Now let us introduce another conjugate pair of free fields. For this we first rename the $(\beta,\gamma)$ system of previous sections as $(\beta_2,\gamma_{-1})$ where the subscripts indicate the conformal dimensions. Next we introduce a new pair $(\beta_3,\gamma_{-2})$ of conformal dimensions 3 and $-2$ respectively. In terms of these two pairs of bosonic ghosts, we attempt to represent the algebra by starting with the standard energy-momentum tensor $T(z)$ for the canonical pairs and choosing $M(z)=\beta_2(z)$. Twisting $T(z)$ by $-a\,\del^3\gamma_{-1}$ reproduces the BMS$_3$ algebra as before. We now need to fix the remaining generators. 

It is easy to imagine that the spin-3 analogue of $M(z)$, namely $V(z)$, should be represented by the spin-3 free field $\beta_3$ up to a normalisation. This takes care of its behaviour as a primary as well as the fact that its OPE with itself is non-singular. The normalisation is not determined by either of these requirements so we leave it free for the moment. We could also twist $T(z)$ by $\del^4\,\gamma_{-2}$, but this would introduce a central term into $T(z)V(w)$ which is not present in the algebra, so we do not carry out this twist. 

The last step is to define $W(z)$. The possible linear terms are $\beta_3$ as well as derivatives of $\gamma_{-1}$ and $\gamma_{-2}$. The derivative terms threaten to render $W$ non-primary, though in the algebra it is expected to be primary. Thus, as in the supersymmetric case, we have to ensure a delicate cancellation of non-primary terms if we introduce them. It will turn out that we only need $\del^5\gamma_{-2}$. Already we face a problem because the cross term between $\beta_3$ and $\del^5\gamma_{-2}$ would give an unwanted sixth-order pole and we need other terms to cancel it. Proceeding further, we add bilinear terms with the restriction that canonically conjugate fields ($\beta_2$ and $\gamma_{-1}$, or $\beta_{3}$ and $\gamma_{-2}$) do not appear in the same term. This allows for bilinears involving $\beta_2$ and $\gamma_{-2}$ together with three derivatives, or bilinears involving $\beta_3$ and $\gamma_{-1}$ with a single derivative. The former case offers four terms with different distributions of derivatives, while the latter case offers two. Thus we introduce six bilinear terms along with all possible linear terms, all with arbitrary coefficients. However it turns out that the system  is overdetermined and one cannot simultaneously satisfy the $T-W, M-W, V-W$ and $W-W$ OPE's. We are therefore led to introduce cubic terms. The computation steadily becomes more complicated as each new term, besides potentially giving a desired term in the OPE, can create a large number of unwanted terms (for example in $W-W$ a new term potentially introduces nine cross-terms!). Despite this, we were able to find precisely two cubic terms, involving two factors of $\beta_2$, one factor of $\gamma_{-2}$ and one derivative. Nested normal-ordering among these has to be carefully prescribed. While it may seem surprising that normal-ordering is relevant at all, given that $\beta_2$ and $\gamma_{-2}$ are mutually commuting fields, this has to do with lower-order poles and becomes very important when the leading poles are of a high order. A related context is the Wakimoto representation of SU(2)$_k$, which crucially involves normal-ordering between mutually commuting fields (see for example Eq.(\ref{eq:1}) of the following section, or Eqs.(15.279) and (15.281) of Ref.\cite{DiFrancesco:1997nk}).

Following the above process we finally arrive at a free-field realisation of all the generators of the $W_3$ BMS$_3$ algebra:
\begin{equation}
  \label{eq:25}
  \begin{split}
    T(z) &=
    -2:\!\beta_2\partial\gamma_{-1}\!:-:\!\partial\beta_2\gamma_{-1}\!:
    - 3:\!\beta_3\partial\gamma_{-2}\!:
    -2:\!\partial\beta_3\gamma_{-2}\!:-a\,\partial^3\gamma_{-1}\\
       W(z) &=\frac{1}{\sqrt{15}}\bigg[
      3:\!\beta_3\partial\gamma_{-1}\!:+:\!\partial\beta_3\gamma_{-1}\!: +5 
    :\!\beta_2\partial^3\gamma_{-2}\!: 
    + :\!\partial^3\beta_2\gamma_{-2}\!: +
    \frac92 :\!\partial^2\beta_2\partial\gamma_{-2}\!: \\ &\qquad\qquad +
    \sfrac{15}{2} :\!\partial\beta_2\partial^2\gamma_{-2}\!: +\frac{8}{a}\Big(:\!\beta_2(:\!\beta_2\partial\gamma_{-2} \!:)\!:+
    :\!\beta_2(:\!\partial\beta_2\gamma_{-2}\!:)\!:\Big)\\
&\qquad\qquad    +\frac{a}{2}\partial^5\gamma_{-2}
      + \frac{68}{15a} \beta_3\bigg]\\
    M(z) &= \beta_2,\qquad    V(z) = -\frac{1}{\sqrt{15}}\beta_3
  \end{split}
\end{equation}
Note the presence of nested normal-ordered products in $W(z)$. These will generate non-trivial contributions to the linear terms in $M$ that are crucial to obtaining the algebra with the right coefficients. One also needs to define the composite fields:
\be
\begin{split} 
\Lambda (w)&=\ :\!TM\!:\!(w)- \frac{3}{10}\partial^2 M(w)\\
\Theta (w)&=\ :\!MM\!:\!(w) 
\end{split}
\ee
 A lengthy and tedious computation results in the following
 non-vanishing OPEs\footnote{We have used the Mathematica package
   Lambda\cite{Ekstrand:2010bp} to compute these OPEs.  This package was
 also used for ensuring proper normal ordering of composite
 operators.}:
  \begin{align}
    T(z)T(w) &~\sim~ \frac{50}{(z-w)^4} + \frac{2T(w)}{(z-w)^2} +
    \frac{\partial T(w)}{(z-w)}\nn\\
    T(z)M(w) &~\sim~ \frac{6a}{(z-w)^4} + \frac{2M(w)}{(z-w)^2} +
    \frac{\partial M(w)}{(z-w)}\nn\\
    T(z)W(w) &~\sim~ \frac{3W(w)}{(z-w)^2} +
    \frac{\partial W(w)}{(z-w)}\nn\\
    T(z)V(w) &~\sim~\frac{3V(w)}{(z-w)^2} +
    \frac{\partial V(w)}{(z-w)}\nn\\
    M(z)W(w) &~\sim~ \frac{3V(w)}{(z-w)^2} +
    \frac{\partial V(w)}{(z-w)}\nn\\
    W(z)W(w) &~\sim~ \frac{100}{3(z-w)^6} + \frac{2 T(w)}{(z-w)^4}
+ \frac{\partial T(w)}{(z-w)^3} \nn\\[1mm]
&\qquad
+ \frac{1}{(z-w)^2}\left[\frac{2}{60}\left(\frac{16}{a}\Lambda- \frac{1088}{15 a^2}
  \Theta\right)+
\frac{3}{10}\partial^2 T\right]\!(w)\nn\\[1mm]
&\qquad+ \frac{1}{(z-w)}\left[\frac{1}{60}\left(\frac{16}{a}\partial\Lambda- \frac{1088}{15 a^2}
  \partial\Theta\right)+
\frac{1}{15}\partial^3 T\right]\!(w)\nn\\[2mm]
    W(z)V(w) &~\sim~ \frac{4a}{(z-w)^6} +\frac{2 M(w)}{(z-w)^4} 
+\frac{\partial M(w)}{(z-w)^3}\nn\\[1mm]
&\qquad+ \frac{1}{(z-w)^2}\left[\frac{2}{60}\left(\frac{16}{a}\Theta\right) +
\frac{3}{10}\partial^2 T\right]\!(w)\nn\\[1mm]
&\qquad+ \frac{1}{(z-w)}\left[\frac{16}{60 a}\partial\Theta+
\frac{1}{15}\partial^3 T\right]\!(w)
\label{eq:13}
  \end{align}
When converted to modes, this precisely agrees with Eq.(\ref{eq:24}). Thus we have obtained a free-field realisation of the $W_3$ BMS$_3$ algebra with central charges in terms of just two canonically conjugate pairs of free fields, $(\beta_2,\gamma_{-1})$ and $(\beta_3,\gamma_{-2})$.

We see that the central charges are $c_1= 100$ and $c_2= 12 a$. Notice that $c_1$ is precisely the critical central charge of the $W_3$ algebra\cite{Bouwknegt:1992wg}. Unlike the previous cases, it would be rather non-trivial to extend the above construction to allow
an arbitrary value for the central charge $c_1$ (including the value $c_1=0$) and we leave this for the future.

\section{An SU(2) generalisation of BMS$_3$ and its Wakimoto representation}

In this section we will use affine current algebra
symmetry to try and derive a generalisation of the BMS$_3$
algebra which one can call ``SU(2)-BMS$_3$''. It contains both the BMS$_3$ algebra and the SU(2) affine Lie algebra as subalgebras. Thereafter we will use the Wakimoto free-field representation\cite{Wakimoto:1986gf} as well as the method of the previous section to find a free-field representation of the full SU(2)-BMS$_3$ algebra. We will be partially successful in this: we do find a realisation of the extended algebra without a Virasoro central charges, but when we incorporate our twist in $T(z)$ to incorporate this central charge, we find other non-central terms in the algebra. Nevertheless, our algebra contains BMS$_3$ and affine SU(2) as subalgebras and each one has the desired central extensions.

Affine symmetry arises from the mode
expansion of conserved currents in a conformal field theory:
\begin{equation}
  \label{eq:8}
  J^a(z) = \sum_{n=-\infty}^{\infty} J^a_n z^{-n-1}, \qquad\quad \bar
  J^a(\bar z) = \sum_{n=-\infty}^{\infty}\bar J^a_n \bar z^{-n-1} ,
\end{equation}
where the index $a$ runs over the dimension $D$ of the Lie algebra.  The
energy-momentum tensor is built out of these currents using the Sugawara
construction(see for example \cite{DiFrancesco:1997nk}): 
\begin{equation}
  \label{eq:19}
  T_J(z)=\frac{1}{2(k+g)}\!:\! J^a(z) J^a(z)\!:\ ,
\end{equation}
where $k$ is the level of the affine algebra and $g$ is the dual
Coxeter number.  The currents transform as dimension-1 primary fields
of the Virasoro symmetry.  This is summarised in the following singular
operator product expansions:
\begin{equation}
  \label{eq:20}
  \begin{split}
    T_J(z)T_J(w)&\sim \half\frac{c}{(z-w)^4}+ \frac{2\,T_J(w)}{(z-w)^2}
    + \frac{\del T_J(w)}{z-w} \\
    J^a(z)J^b(w)&\sim \frac{k\delta^{ab}}{(z-w)^2}+
    \frac{if^{abc}J^c(w)}{z-w} \\
    T_J(z)J^a(w) &\sim \frac{J^a(w)}{(z-w)^2}+\frac{\del J^a(w)}{z-w} ,
  \end{split}  
\end{equation}
where the Virasoro central charge $c= kD/(k+g)$ and $f^{abc}$ are the
structure constants of the Lie algebra.

Let us specialise to SU(2). In our conventions this algebra has $f^{abc}=\sqrt2\,\epsilon^{abc}$. Using the basis $J^\pm=\frac{1}{\sqrt2}(J^1\pm iJ^2),J^0=\sqrt2\, J^3$, the current algebra is:
\begin{equation}
  \label{eq:2}
  \begin{split}
    J^+(z)J^-(w) &\sim \frac{k}{(z-w)^2} + \frac{J^0(w)}{z-w} , \\
    J^0(z)J^\pm(w) &\sim \frac{\pm 2 J^\pm}{z-w} ,\\
    J^0(z)J^0(w) &\sim \frac{2k}{(z-w)^2} .
  \end{split}
\end{equation}
The remaining products are all regular, in particular $J^+(z)J^+(w)$
has no singularity. We will make use of this fact later on.

Next, the Sugawara construction in this basis looks like:
\be
 T_J(z) = \frac{1}{2(k+2)}\left[\frac12 :\!J^0J^0\!:\!(z) + :\!J^+J^-\!:\!(z) +
    :\!J^-J^+\!:\!(z)\right] 
\ee

In order to combine this with BMS$_3$, we actually need a variant of the above construction. In fact it has long been known that there is a twisted version of the SU(2) current algebra, first introduced in Ref.\cite{Knizhnik:1988ak} in the context of 2d gravity, in which $(J^+,J^0,J^-)$ have conformal dimensions $(2,1,0)$ respectively. This version is obtained by modifying the above Sugawara energy-momentum tensor as:
\be
T(z)= T_J(z)-\half \del J^0(z)
\ee
It is easy to verify that after this twist, the conformal dimensions of the currents change as above. We now have a potential way of defining a combined SU(2)-BMS$_3$ algebra. Note that $T(z)$ and $J^+(z)$ together form a pair of spin-2 holomorphic fields of which the former satisfies a Virasoro algebra, the latter has a non-singular OPE with itself and the latter is a spin-2  primary under the former. These are all the ingredients that define a BMS$_3$ algebra with a vanishing second central charge $c_2$. This leads us to define the SU(2)-BMS$_3$ algebra with central extensions by introducing a $c_2$ term in the $T-J^+$ OPE, leading to:
\begin{equation}
  \begin{split}
T(z)T(w)&\sim \half\frac{c_1}{(z-w)^4}+ \frac{2T(w)}{(z-w)^2}+
\frac{\del T(w)}{z-w}\\ 
T(z)J^+(w)&\sim \half\frac{c_2}{(z-w)^4}+ \frac{2J^+(w)}{(z-w)^2}+\frac{\del J^+(w)}{z-w}\\
J^+(z) J^+(w)&\sim 0,\\
 J^+(z)J^-(w) &\sim \frac{k}{(z-w)^2} + \frac{J^0(w)}{z-w} , \\
    J^0(z)J^\pm(w) &\sim \frac{\pm 2 J^\pm}{z-w} ,\\
    J^0(z)J^0(w) &\sim \frac{2k}{(z-w)^2},\\
T(z)J^0(w) &\sim \frac{2k}{(z-w)^3} + \frac{J^0(w)}{(z-w)^2}+\frac{\del J^0(w)}{z-w}\\
T(z)J^-(w)&\sim \frac{\del J^-(w)}{z-w}
\end{split}
\label{su2-BMS}
\end{equation}
At this stage it is not clear what central terms, or other terms, we need to incorporate in the last two OPE's to make a consistent algebra. While in previous sections of this paper the algebra including central extensions was completely specified in advance, here we will take a different approach -- to let the free-field realisation determine what additional terms the algebra should contain.

Let us now introduce the Wakimoto representation of the SU(2) affine Lie algebra\cite{Wakimoto:1986gf} (see also Ref.\cite{DiFrancesco:1997nk}). Recall that this has been one of the most important uses of $(\beta,\gamma)$ systems  and enables us to construct the affine SU(2) algebra at arbitrary level $k$. In this representation, the three holomorphic SU(2) currents $J^\pm(z), J^0(z)$ are constructed out of $\beta,\gamma$ and a free scalar $\varphi$ with a background charge that depends on a real number $k$. For this purpose, the $(\beta,\gamma)$ pair is taken to have spin $(1,0)$. The precise construction is as follows:
\begin{equation}
  \label{eq:1}
  \begin{split}
    J^+(z) &= \beta(z), \\
    J^0(z) &= \frac{i\sqrt{2}}{\alpha_+}\partial\varphi(z) +
    2:\!\gamma\beta\!:\!(z), \\
    J^-(z) &=\frac{-i\sqrt{2}}{\alpha_+}:\!\partial\varphi\gamma\!:\!(z) -
    k\partial\gamma(z) - :\!\beta\gamma\gamma\!:\!(z) \ .
  \end{split}
\end{equation}
Here $k$ is the level of the affine algebra and
$\alpha_+=\frac{1}{\sqrt{k+2}}$ is proportional to the background
charge of the scalar field $\varphi$.  Via the canonical OPE's of $\beta$ with $\gamma$ and $\del\varphi$ with itself, one can show that the above spin-1 currents satisfy the affine SU(2) algebra at arbitrary level $k$:

The energy-momentum tensor in this representation is given by:
\begin{equation}
  \label{eq:3}
  \begin{split}
    T_J(z) &= \frac{1}{2(k+2)}\left[\frac12 :\!J^0J^0\!:\!(z) + :\!J^+J^-\!:\!(z) +
    :\!J^-J^+\!:\!(z)\right] \\
    & = -:\!\beta\partial\gamma\!:\!(z) -\frac12
    :\!\partial\varphi\partial\varphi\!:\!(z) -
    \frac{i\alpha_+}{\sqrt{2}}\partial^2\varphi(z)\ .
  \end{split}
\end{equation}
We see that in the Wakimoto form it splits into two pieces
corresponding to the energy-momentum tensor of a $\beta$-$\gamma$
system and that of a scalar field $\varphi$ with a background
charge.  Comparing with the canonical form for the latter, we find that
the background charge for $\varphi$ is $-\alpha_+/2$. By virtue of
being one of the currents in the Wakimoto representation it is obvious
that the conformal dimension of $\beta$ is $1$. It 
follows that its canonically conjugate field $\gamma$ has
conformal dimension $0$.

Next we ask how, in the Wakimoto representation, one can perform a twist to change the spins of $(J^+,J^0,J^-)$ to $(2,1,0)$. This has already been carried out in Ref.\cite{Mukhi:1993zb} where it was shown that the above twist can be implemented within the Wakimoto representation precisely by changing the spins of $(\beta,\gamma)$ to $(2,-1)$. Therefore to realise SU(2)-BMS$_3$ with an arbitrary central charge, we must perform this twist of the Wakimoto representation, and then twist the energy-momentum tensor of the theory by $\del^3\gamma$ as in the previous section. We have already shown that in this situation $T(z)$ and $\beta(z)$ satisfy a BMS$_3$ algebra. The new ingredient is that we now have additional generators $J^0,J^-$ that extend BMS$_3$ and we will find that the OPE's come out as specified in Eq.(\ref{su2-BMS}) plus some additional terms.

The first twist gives, in the Wakimoto representation:
\be
\begin{split}
{\tilde T}_J(z)&=T_J(z)-\del J^0(z)\\
& = -2:\!\beta\partial\gamma\!: - :\!\gamma\partial\beta\!: -\frac12 :\!\partial\varphi\partial\varphi\!:
  -\frac{i}{\sqrt{2}} \left(\alpha_++\frac{1}{\alpha_+}\right)\partial^2\varphi .
\end{split}
\ee
With respect to $\tT$, the
$\beta$-$\gamma$ system now has conformal dimension $(2,-1)$ and as desired, this changes the
conformal dimension of the $J^+$ current to 2 and
that of $J^-$ to 0.  Since $J^+=\beta$ has a trivial OPE with itself,
we get the following set of OPE's:
\begin{equation}
  \label{eq:4}
  \begin{split}
T(z)T(w) &\sim \half\frac{c_1}{(z-w)^4}+\frac{2T(w)}{(z-w)^2}+\frac{\del t(w)}{z-w},\\
 T(z)J^+(w)&\sim \frac{J^+(w)}{(z-w)^2}+\frac{\del J^+(w)}{z-w},\\
    J^+(z)J^+(w)&\sim 0 ,
  \end{split}
\end{equation}
where $c_1 = 27 - 6\left(\alpha_++\frac{1}{\alpha_+}\right)^2$ is the
Virasoro central charge.  Since $J^+$ is a Virasoro primary, we do
not find a central extension term in the second commutator. This
limitation can be overcome by further deforming the energy-momentum
tensor, defining $T = \tilde{T} - a\, \partial^3\gamma$.  Now the
operator product expansion of $T$ with $J^+$ has a fourth order pole
which gives rise to the desired central extension. Just as in the
previous section, this deformation does not produce any unwanted terms
in the Virasoro algebra or change its central charge $c_1$. 

The final energy-momentum tensor then takes the form:
\begin{equation}
  \label{eq:5}
T = -2:\!\beta\partial\gamma\!: - :\!\gamma\partial\beta\!: -\,
  a:\!\partial^3\gamma\!: -\frac12 :\!\partial\varphi\partial\varphi\!:
  -\frac{i}{\sqrt{2}} \left(\alpha_++\frac{1}{\alpha_+}\right)\partial^2\varphi .
\end{equation}
The OPEs between $T(z)$ and $J^+(z)$
give the BMS$_3$ algebra, namely, we identify $M(z) = J^+(z)$ and
the resulting OPEs are as in Eq.(\ref{eq:4}) except that the middle line is modified to:
\begin{equation}
  \label{eq:6}
    T(z)J^+(w) \ \sim\  \half \frac{12a}{(z-w)^4} +
  \frac{2J^+(w)}{(z-w)^2}+\frac{\del J^+(w)}{z-w}
\end{equation}
We see that $c_2=12a$.  But now this structure is supplemented by the remaining SU(2)
currents. The additional OPEs between $T$ and these currents are:
\begin{equation}
  \label{eq:26}
  \begin{split}
    T(z)J^0(w) &\ \sim \ \frac{12a \gamma(w)}{(z-w)^4}
      +\frac{2k}{(z-w)^3} +\frac{J^0(w)}{(z-w)^2} + \frac{\partial
        J^0(w)}{z-w}, \\
      T(z)J^-(w) &\ \sim \ \frac{-6 a:\!\gamma\gamma\!:(w)}{(z-w)^4} +
      \frac{\partial J^-(w)}{z-w}.
  \end{split}
\end{equation}

Under the twisting which takes us from $T_J$ to
$\tilde{T}$ the Wakimoto currents $J^\pm$ remain primary while their conformal dimensions shift from $(1,1)$ to
$(2,  0)$. Meanwhile the current $J^0$ is no longer primary, acquiring a central charge proportional to $k$ after the twist. However when we go from
$\tilde{T}$ to $T$, none of the currents remains a primary field.  The
operator product expansion of $T$ with $J^0$ acquires a central term
proportional to $k$ with a cubic pole, as well as a term proportional to $a$ multiplying the free field $\gamma$ with a quartic pole. For $T$ with $J^-$, one finds a non-central fourth-order pole. However since the
modification of the energy-momentum tensor does not affect the current
algebra as such, the system continues to have SU(2) affine symmetry. So it may be interesting as a non-Abelian extension of BMS$_3$. As in the previous section, here too we can couple this system to a
conformal field theory of chiral matter with central charge $c_0$,
which allows us to have arbitrary value for the central charge $c_1$.

\section{Conclusions and Future Directions}

We have shown that the BMS$_3$ algebra with arbitrary central charges $c_1,c_2$ can be written in terms of a twisted chiral conformal field theory of free fields coupled to
arbitrary chiral matter.  We also constructed a free-field realisation for the supersymmetric extension of the
BMS$_3$ algebra and the higher-spin $W_3$-BMS$_3$ algebra. Finally, we discussed a coupled SU(2)-BMS$_3$
system using the Wakimoto free field representation of the chiral
SU(2) affine algebra. In the BMS$_3$ and supersymmetric cases, we were able to tune the central charge $c_1$ to zero if desired by adding suitably (non-unitary, in the usual sense) matter. For $W_3$ we did not yet show how to tune $c_1$ to an arbitrary value. It is important to note that $c_2\ne 0$ in our construction, which is achieved by a twist of the energy-momentum tensor and its generalisations. It remains an interesting open question to generalise our construction to the $W_n$ ($n>3$) and $W_\infty$ versions of BMS$_3$. 

Our work suggests  a variety of possible directions to pursue. First of all, one should understand whether there is any way to restore unitarity, which as it stands is broken by the highest-weight conditions. One may then ask what is the bulk dual for our theories and what quantity in the bulk is computed by correlation functions of our free-field realisation.  Beyond the fact that the bulk theory contains asymptotically flat gravity, it may also contain various other types of fields and we have yet to initiate a detailed investigation of this.

Another interesting observation is that these chiral CFT-like representations
suggest possible relations between the asymptotic BMS$_3$ symmetry
algebra and the symmetry algebra of open string theory. If we
choose a conformal field theory with central charge $c_0=0$ (in the
$\beta$-$\gamma$) formulation or $c=6(\alpha_+ + \frac{1}{\alpha_+})^2 -1$ (in
the Wakimoto $SU(2)$ theory), we can get a critical open string
theory representation with BMS$_3$ symmetry. One possibility is to avoid adding a matter part entirely, but instead to add a pair of $(b,c)$ ghosts of spins $(2,-1)$ (to be viewed as worldsheet ghosts, and not to be confused with the supersymmetric $(b,c)$ system discussed above). Passing to the BRST cohomology, we then appear to find a topological string with BMS$_3$ symmetry. It would be interesting to investigate this possibility.



\section*{Acknowledgements}

We are grateful to Rudranil Basu for helpful discussions. 
We thank IISER, Mohali for hospitality during completion
of this work. DPJ would like to thank IISER, Pune for hospitality
during the course of the work.  Our work is partially supported by the following Government of India Fellowships/Grants: NB by a Ramanujan Fellowship, DST; DPJ by DAE XII-plan grant 12-R\&D-HRI-5.02-0303; SM by a J.C. Bose
Fellowship, DST; and TN by a UGC Fellowship. We thank the people of India for their generous
support for the basic sciences.

\bibliographystyle{JHEP}
\bibliography{bms3}

\end{document}